\documentclass{article}
\usepackage{spconf,amsmath,graphicx}
\usepackage{multirow,booktabs,url}
\usepackage[T1]{fontenc}
\usepackage{color,soul}
\usepackage{hyperref}
\usepackage{mathtools}
\usepackage{pifont}
\usepackage{amsfonts}
\usepackage[inline]{enumitem}

\newcommand{\cmark}{\ding{51}}
\newcommand{\xmark}{\ding{55}}

\DeclareMathOperator{\bernoulli}{Bernoulli}

\title{CarneliNet: Neural Mixture Model for automatic speech recognition}

\name{Aleksei Kalinov{\normalfont \textsuperscript{1}}\sthanks{Work was done during an internship at NVIDIA},
Somshubra Majumdar, Jagadeesh Balam, Boris Ginsburg
}
\address{NVIDIA, USA\\
\textsuperscript{1}Skolkovo Institute of Science and Technology, Russia
}

\begin{document}
\maketitle

\begin{abstract}
End-to-end automatic speech recognition systems have achieved great accuracy by using deeper and deeper models. 
However, the increased depth comes with a larger receptive field that can negatively impact model performance in  streaming scenarios. We propose an alternative approach that we call Neural Mixture Model. The basic idea is to introduce a parallel mixture of shallow networks instead of a very deep network. To validate this idea we design CarneliNet -- a CTC-based neural network composed of three mega-blocks. Each mega-block consists of multiple parallel shallow sub-networks based on 1D depthwise-separable convolutions. We evaluate the model on LibriSpeech, MLS and AISHELL\nobreakdash-\nobreakdash2 datasets and achieved close to state-of-the-art results for CTC-based models. Finally, we demonstrate that one can dynamically reconfigure the number of parallel sub-network to accommodate the computational requirements without retraining.

\end{abstract}


\section{Introduction}
\label{sec:intro}
End-to-end automatic speech recognition (ASR) systems based on neural networks (NN) have become a new state-of-the-art. While the first NN-based ASR models (2014, \cite{DeepSpeech2014} and \cite{Graves2014}) had only 5-6 hidden layers, the latest  ASR models are much deeper. Collobert et al \cite{collobert2016} proposed Wav2Letter, a fully-convolutional  model with 12 layers to allow the network to "see" a larger context.  
Jasper \cite{li2019jasper} added to the model residual connections, which allowed increasing the depth to 54 convolutional layers. 
QuartzNet~\cite{kriman2020quartznet} and TDS~\cite{hannun2019} models replaced regular convolution with depth-wise separable convolutions to make the model deeper while keeping model size small. For example QuartzNet-15x5 has 74 convolutional layers and a very large receptive field.  

Dong et al \cite{Dong2018Transformer} adapted a Transformer model to ASR. The Transformer approach was further expanded in \cite{Tsunoo2019, Mohamed2019Transformers, tsunoo2019online, yeh2019transformertransducer}. One of the latest model in the attention-based line, Conformer ~\cite{gulati2020conformer}, is an RNN-Transducer with a very deep encoder composed of 16 stacked residual blocks with each block consisting of two feed-forward blocks, a convolution  module and multi-headed self-attention module stacked together. Han et al proposed ContextNet~\cite{han2020contextnet} - an RNN-Transducer with a global context mechanism based on squeeze-and-excitation. The ContextNet encoder has 23  blocks with 5 depth-wise separable convolutional layers per block. 

The increase in the model depth has a few negative side-effects on model performance, e.g. slower training, higher memory consumption and increased inference latency. In this work, we explore an alternative way to scale networks with width, \textit{without increasing parameter count}. We propose CarneliNet - a non-autoregressive, hierarchical, Neural Mixture Model trained using Connectionist Temporal Classification (CTC) loss. The model consists of a stack of 3 mega-blocks, where each mega-block is comprised  of a large number of relatively shallow parallel convolutional modules. These sub-blocks, which we call \textit{towers}, are independent, so they can be computed in parallel to reduce inference latency.  
We further leverage a dropout technique on these parallel blocks, which we refer to as \textit{tower dropout}, to strengthen the independence of each block inside a mega-block from other "parallel" blocks. We demonstrate that this augmentation technique allows us to scale the network's computational requirements during inference \textit{without retraining the model}.
The model also has a relatively small receptive field which makes it a good candidate for streaming ASR.

\begin{figure*}[ht!]
 \centering
  \includegraphics[width=0.95\linewidth]{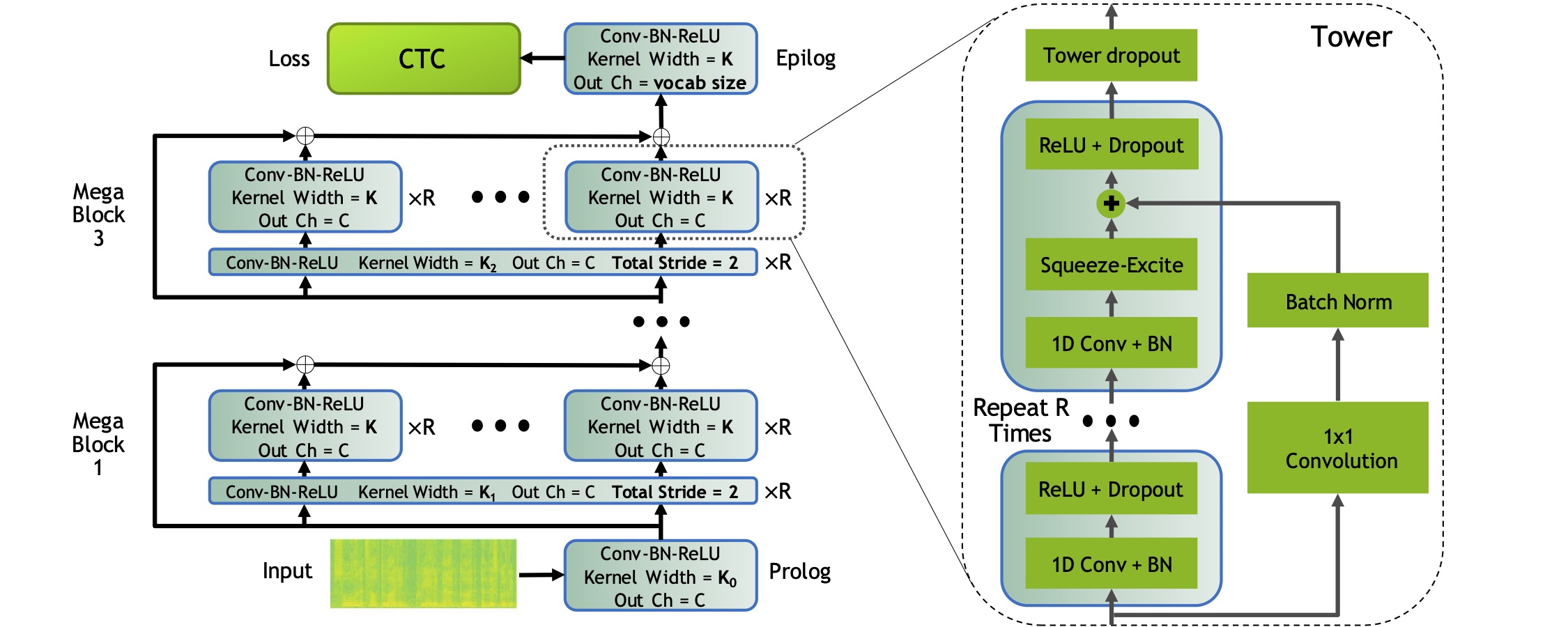}
 \caption{Architecture of CarneliNet-$C$. The model starts with a prologue block, followed by 3 mega-blocks with, respectively, 5, 6, and 7 parallel towers, and an epilogue block. Each mega-block starts with a 2x-downsampling operation and then branches out into several independent towers with $R$ repeated residual blocks. Each tower has exactly $C$ channels. Towers finish with a \textit{tower dropout} operation which allows us to remove towers during inference without significant harm to the accuracy. Tower outputs are summed together. } 
 \label{fig:carnelinet_arch}
\end{figure*}

\section{Related work}

Recent work on multi-stream CNNs \cite{han2019mutlistream} introduces a model that uses several independent multi-resolution time-delay NNs (TDNNs). The TDNN-F model consists of  parallel streams of self-attention encoders, where each stream has 3 layers of 1D convolutions with dilated kernels whose dilation rates  are unique for a given stream, followed by a self-attention layer and factorized feed-forward layer. The final embedding layer concatenates the encoder output from each stream and linearly projects the concatenated vector to the final embedding. In the follow-up paper \cite{han2021mutlistream}, the authors explored a deeper version of multi-stream TDNN models with a large numbers of streams. A similar approach was suggested in \cite{luo2020multiquartznet} which replaces the original QuartzNet 1D time-channel separable convolution with multi-stream convolutions. Each stream has a unique dilated stride on convolutional operations. Streams are combined through channel-wise attention module, which calculates the attention weight of each convolutional stream by spatial channel-wise pooling.


While our idea resembles this prior work, we emphasise a few notable differences: information sharing between parallel feature extractors via aggregation at the end of mega-blocks instead of fully independent streams, uniform time-channel separable convolutions with a single kernel size instead of dilated convolutions. This work also presents an analysis of the impact of individual towers on the final result. Additionally, the absence of large kernels and dilated convolutions results in a smaller receptive field than other multi-stream approaches, making the model more suitable for streaming ASR.


\section{Model architecture}
\label{sec:architecture}
CarneliNet is a convolutional CTC model with several parallel towers processing the same input and combining outputs at each stage. 
Each tower has a Citrinet-like structure \cite{citrinet} consisting of repeated 1D time-channel separable convolutions with 1D Squeeze-and-Excitation (SE) \cite{hu2018squeeze} context modules.
We do not experiment with the internals of these computational blocks, instead focusing on analysis of their composition with respect to each other. In this light, CarneliNet derives from Citrinet \cite{citrinet}, QuartzNet \cite{kriman2020quartznet} and ContextNet \cite{han2020contextnet} and reuses their building blocks in a unique way.

Compositionally, the model has 5 parts: a prologue block, 3 mega-blocks with, respectively, 5, 6 and 7 parallel towers, and an epilogue block. 
All mega-blocks start with a block that consists of several 1D time-channel separable convolutions, the last of which has stride 2. The output of the down-sampling operation is passed as an input to several towers that compute outputs independently from each other, in parallel. Each tower consists of $R$ repeated QuartzNet blocks \cite{kriman2020quartznet} with the same kernel size and finishes with an SE module, similar to ContextNet \cite{han2020contextnet}. The number of channels in each tower, denoted as $C$, is kept constant for simplicity. As this number directly controls computational power of the model as demonstrated in \autoref{sec:ablation_study}, we refer to different variants of the models as \textit{CarneliNet-C}. The schematic of the \textit{CarneliNet-C} architecture is depicted in \autoref{fig:carnelinet_arch}.

In order to further simplify the design of the network, we choose to keep kernel size the same across all towers. While it is possible to tune the kernel size of each individual tower and gain an small improvement in word error rate (WER), finding an exact configuration of kernel sizes requires a significant number of experiments and computation. Additionally, we aim to keep the receptive field small while maintaining comparable speech recognition quality by favouring smaller kernel sizes. Based on the results of ablations in section \ref{Kernel_configurations}, we choose $11$ as a primary kernel size.

Tower outputs are added together to obtain a single output of the mega-block. While different aggregation strategies are possible, we use a summation, shifting computation to the towers themselves. We found that summation worked no worse than a family of weighted per-output or per-channel summations with or without dependence on the input. 

Independent parallel computation of towers gives us the flexibility to choose which outputs are going to contribute to the final result. To save computational resources we can reduce the number of towers in any mega-block by discarding a few of them from the computational graph after the full model is trained. This allows us to dynamically change the computational complexity of the model without additional training as we show in section \ref{sec:Tower_dropout}.

\subsection{Tower Dropout}

To overcome the inevitable reduction in quality caused by removal of towers during inference, we introduce \textit{tower dropout}. The idea of tower dropout is similar to Stochastic Depth \cite{huang2016deep} but dropout is applied to outputs of towers before final summation in the mega-block. Dropping an output of a tower before summation ignores the computations of all blocks in the tower, effectively reducing the model width. 

More formally, if we denote the computed tensor at the output of tower $i$ as $O_i \in \mathbb{R}^{T \times C}$, where $T$ is number of output timestamps and $C$ is the number of channels, then during training time the final output $M \in \mathbb{R}^{T \times C}$ of a mega-block  with $N$ towers is computed as 
$M = \sum_{i=1}^{N} w_i O_i / p$, where $w_i \sim \bernoulli(p)$ is a Bernoulli random variable with probability $p$ of being $1$. We use this technique of scaling outputs by inverse survival probability in order to avoid the rescaling during the inference step.

During inference we select the active towers by setting a respective mask variable $\delta_i$ to $1$. Setting the variable $\delta_i$ to $0$ ignores the output of $i$-th tower. Using these masks we obtain the final weights $w_i = \delta_i / K$, where $K = \sum_{i=1}^{N} \delta_i$ is the number of kept towers. Final weights are used directly to compute an inference output of mega-block $M = \sum_{i=1}^{N} w_i O_i$. The re-scaling of mask variables is necessary to keep the expected value of the output unchanged $\mathbb{E}\left[M\right] = \sum_{i=1}^{N} w_i \mathbb{E}\left[O_i\right] = \sum_{i=1}^{N}  \frac{\delta_i}{K} \mathbb{E}\left[O_i\right] = \mathbb{E}\left[O\right] \frac{1}{K} \sum_{i=1}^{N}  \delta_i = \mathbb{E}\left[O\right]$, assuming the expected value of each tower output $\mathbb{E}\left[O_i\right]$ is the same and denoted as $\mathbb{E}\left[O\right]$.

\section{Experiments}

\subsection{LibriSpeech}
\label{sec:librispeech}

We train a number of CarneliNets with various numbers of channels per tower on the LibriSpeech (LS) dataset~\cite{panayotov2015librispeech}. 
The LS training set was used to train a word-piece tokenizer with 1024 subwords with the help of the Huggingface library~\cite{wolf2020huggingface}.

We used the NovoGrad optimizer \cite{novograd2019} to train all the models. Experimentally we found that a high initial learning rate (LR) of $0.1$ works best for this relatively shallow model. The other parameters of the optimizer were set to $\beta_1=0.8, \beta_2=0.25$, similar to Citrinet \cite{citrinet}.
We utilize a cosine annealing scheduler \cite{loshchilov2016sgdr} with a final LR of $10^{-5}$ preceded by 1000 steps of warmup during training. We also use dropout \cite{srivastava14a} with value $0.1$ for all the models.
SpecAugment \cite{park2019} was used as the main data augmentation method with $2$ frequency masks for all models, $2$ time masks for CarneliNet-256 and CarneliNet-384, and $10$ time masks for larger models.
The training of models lasted $1000$ epochs and was performed on 32 V100 GPUs with a batch size of 32.

We also use two external language models (6-gram and transformer) to perform beam rescoring.  These models were trained on the LibriSpeech language model training text combined with train transcripts. 

The results are presented in Table~\ref{tab:LibriSpeech}. We compare them with two CTC-based models Citrinet \cite{citrinet}, a deep model from which CarneliNet's computational sub-networks were derived, and Multistream CNN \cite{han2021mutlistream}, a recent model that employs parallel streams of self-attention encoders with different dilation rates. We also provide results for RNN-transducer \cite{graves2012transducer} models, namely ContextNet \cite{han2020contextnet} and Conformer \cite{gulati2020conformer} for reference. We demonstrate that CarneliNet achieves scores close to the state-of-the-art non-autoregressive CTC-based models while being highly scalable.  

\begin{table}[thb!]
\centering
\caption{CarneliNet vs state-of-the-art CTC-based and RNNT-transducer-based models, LibriSpeech, WER($\%$)}
\vspace{4pt}
\label{tab:LibriSpeech}
\scalebox{0.90}{
\begin{tabular}{cccccc} 
 \toprule
 \multirow{2}{*}{\textbf{Model}}    &
 \multirow{2}{*}{\textbf{LM}}       &
 \multicolumn{2}{c}{\textbf{Test}}  &
 \textbf{Params,}  \\
  &&\textbf{clean}& \textbf{other}& \textbf{M}  \\
\hline
  \multirow{3}{*}{Citrinet-1024 \cite{citrinet}}
    & -      & 2.52 & 6.22 & \multirow{3}{*}{142} \\
    & 6-gram & 2.10  & 5.06 & \\
    & Transf & 2.00  & 4.69 & \\
\hline
  \multirow{3}{*}{Multistream CNN \cite{han2021mutlistream}}
    & -      & 3.52 & 8.32 & \multirow{3}{*}{93.9} \\
    & 4-gram & 2.80  & 7.06 & \\
    & SRU & 1.75  & 4.46 & \\

    \hline
  \multirow{2}{*}{ContextNet \cite{han2020contextnet}}
    & -      & 2.1 & 4.6 & \multirow{3}{*}{112.7} \\
    & LSTM & 1.9  & 4.1 & \\

    \hline
  \multirow{2}{*}{Conformer \cite{gulati2020conformer}}
    & -      & 2.1 & 4.3 & \multirow{3}{*}{118.8} \\
    & LSTM & 1.9  & 3.9 & \\

    \specialrule{\heavyrulewidth}{0.1em}{0.1em}

 \multirow{3}{*}{CarneliNet-256} &  -      & 4.52 & 11.59 & \multirow{3}{*}{10} \\
    & 6-gram  & 3.30 & 8.41 \\
    & Transf  & 2.77 & 7.35 \\
 \midrule
  \multirow{3}{*}{CarneliNet-384} 
    & -      & 3.60  & 9.37  &  \multirow{3}{*}{21.1} \\
    & 6-gram & 2.73 & 7.11 & \\
    &  Transf & 2.48 & 6.34 & \\
 \midrule
  \multirow{3}{*}{CarneliNet-512}
    &  -     & 3.62 & 8.90 & \multirow{3}{*}{36.5} \\
    & 6-gram & 2.81  & 6.82 & \\
    & Transf & 2.51 & 6.06 & \\
 \midrule
  \multirow{3}{*}{CarneliNet-1024}
    & -      & 2.94 & 7.16 & \multirow{3}{*}{142} \\
    & 6-gram & 2.45  & 5.87 & \\
    & Transf & 2.32  & 5.31 & \\
 \bottomrule
\end{tabular}
}
\end{table}

\subsection{MLS}
\label{sec:mls}

We train the largest model, CarneliNet-1024, on the English part of the MLS dataset \cite{MLS} which   includes 44K hours of English read-speech speech. 
Following \cite{citrinet}, we filter out training examples with typos and misalignments in transcripts, utterances in Spanish and Russian from English portion of MLS dev and test sets and transcripts with periods from test set. We also replace hyphens with space characters to keep consistency with the LibriSpeech dataset. 
The model hyper-parameters remained the same as for LibriSpeech with the exception of reducing the learning rate to $0.05$ and using a 1024 sentence piece tokenizer \cite{kudo2018sentencepiece} built on the MLS English training set corpus. The model was trained for 100 epochs on 256 GPUs with batch size 32 per device. We show the results in Table \ref{tab:MLS}. The shallower CarneliNet performs on par with its much deeper counterpart Citrinet.

\begin{table}[thb!]
\centering
\caption{CarneliNet-1024 trained on normalized MLS English, evaluated on normalized MLS English and LibriSpeech-other,  WER($\%$)}
\vspace{4pt}
\label{tab:MLS}
\scalebox{0.85}{
 \begin{tabular}{ccccccc}
\toprule
\multirow{2}{*}{\textbf{Model}}                              & \multirow{2}{*}{\textbf{LM}} & \multicolumn{2}{c}{\textbf{LS-other}} & \multicolumn{2}{c}{\textbf{MLS}} & \textbf{Params,}     \\
                                                    &                     & \textbf{dev}  & \textbf{test} & \textbf{dev}        & \textbf{test }      & \textbf{M}           \\
                                                    \midrule
\multirow{3}{*}{Citrinet-1024\cite{citrinet}}             & -                   & 5.79          & 5.69          &  6.99          &     8.46       & \multirow{3}{*}{142} \\
                                                    & 6-gram              & 4.72          & 4.83          &     5.76       &        6.79    &                      \\
                                                    & Transf              & 4.41          & 4.62          &     5.44       &      6.39      &                      \\ 
                                                    \specialrule{\heavyrulewidth}{0.1em}{0.1em}
\multirow{3}{*}{CarneliNet-1024}                    & -                   & 5.86    	  & 5.73    &   	6.98    &	8.32           & \multirow{3}{*}{142} \\
                                                    & 6-gram              & 4.97          & 5.04          &        5.89    &    7.02        &                      \\
                                                    & Transf              & 4.74          & 4.86          &     5.60       &   6.68         &                      \\
                                                    \bottomrule
                                                
\end{tabular}
}
\end{table}



\subsection{AISHELL-2}
\label{sec:aishell}
We trained a CarneliNet-1024 model on AISHELL-2 \cite{du2018aishell}, a Mandarin corpus with 1000 hours of training speech. Unlike the English language models, we used character level tokenization instead of sentence piece tokenization for Mandarin ASR. The model was trained from scratch using 64 V100 GPUs for 400 epochs with all the other parameters same as the ones used for LibriSpeech in section \ref{sec:librispeech}, except for the initial LR which was set at $0.05$.
For beam rescoring, we used a 4-gram LM trained on the training corpus of AISHELL-2 with the beam width 128.

In Table \ref{tab:aishell} we compare the results with other strong models. CarneliNet produces competitive scores with just greedy decoding compared to the autoregressive Continuous Integrate-and-Fire (CIF) \cite{cif2020} model that relies on frame-by-frame computation of integration weights and with a two pass rescoring U2++ Conformer \cite{u2ppwu2021}. 
Compared to non-autoregressive counterparts, such as NAR-Transformer \cite{NAR2021}, LASO \cite{bai2021fastbert} and QuartzNet, CarneliNet achieves the best CER score on this dataset.

\begin{table}[thb]
\centering
\caption{AISHELL-2: Carnelinet-1024 trained from scratch, CER($\%$)}
\vspace{4pt}
\label{tab:aishell}
\scalebox{0.90}{
\begin{tabular}{ccccc} 
 \toprule
 \textbf{Model} &  LM & \textbf{iOS} &  \textbf{Android} & \textbf{mic} \\
 \midrule
 \multirow{2}{*}{U2++ Conformer\cite{u2ppwu2021}}  & 0-LM & 6.70 & - & - \\
       & Attention & 5.50 & - & - \\
  \midrule 
    CIF \cite{cif2020, dong2020comparison} & SAN & 5.68 & 6.09 & 6.20 \\
 \midrule
  QuartzNet \cite{balam2020improving} &  - & 6.96 & - & - \\
  \midrule
 NAR-Transformer \cite{NAR2021} &  - & 7.1 & 8.0 & 8.1 \\
   \midrule
 LASO-large w/ BERT \cite{bai2021fastbert} &  - & 6.5 & 7.2 & 7.1 \\
 \specialrule{\heavyrulewidth}{0.1em}{0.1em}
 \multirow{2}{*}{Carnelinet-1024} &  - &  6.2 & 6.45 & 6.48 \\
 &  4-LM & 5.8  & 6.07  & 6.05\\
 \bottomrule
\end{tabular}
}
\end{table}
\section{Ablation study}
\label{sec:ablation_study}

We evaluate the contribution of each component of the model to the final accuracy by performing extensive ablation studies. Specifically, we choose CarneliNet-$384$ with kernel size $11$ as a baseline model.
All experiments are performed on the LibriSpeech dataset. Models were trained for $400$ epochs with exactly the same parameters as in \autoref{sec:librispeech}.  The hardware setup is also identical with 32 V100 GPUs, each processing a batch of size 32.

\subsection{Kernel configurations}
\label{Kernel_configurations}

We first evaluate the effect of kernel size on the final quality of the model. We train models with odd kernel sizes from $3$ to $23$ and evaluate the performance on dev and test portions of LibriSpeech datasets. 
The best results, as shown in Table \ref{tab:Ablation_tower_configuration}, are observed with larger kernel sizes as the number of parameters and receptive field naturally grow. However, the improvement in recognition accuracy is negligible beyond kernel size of $11$. Smaller kernels size provides us  with the best trade-off between WER and receptive field size for potential streaming applications.

\begin{table}[thb]
\centering
\caption{Accuracy vs kernel size. CarneliNet-384, LibriSpeech, greedy WER($\%$).}
\vspace{4pt}
\label{tab:Ablation_tower_configuration}
\scalebox{0.90}{
\begin{tabular}{cccccc} 
 \hline
 \textbf{Kernel}    &
 \multicolumn{2}{c}{\textbf{dev}} &
  \multicolumn{2}{c}{\textbf{test}} & 
   {\textbf{Params,}} \\
  \textbf{size} & \textbf{clean} & \textbf{other} & \textbf{clean}& \textbf{other} & \textbf{M}\\
 \hline
    3	&   3.87    &	10.52   &   4.10    &	10.46   &	20.7 \\
    5	&   3.72    &	9.95    &   3.84    &	9.96    &	20.7 \\
    7	&   3.66    &	9.72    &   3.81    &	9.83    &	20.8 \\
    9	&   3.60    &	9.95    &   3.88    &	9.79    &	20.9 \\
    \textbf{11}	& \textbf{3.49} & \textbf{9.56} & \textbf{3.70}  &	\textbf{9.32}& \textbf{21.0} \\
    13	&   3.38    &	9.65    &   3.71    &	9.47    &	21.0 \\
    15	&   3.47    &	9.46    &   3.70    &	9.35    &	21.1 \\
    17	&   3.40    &	9.49    &   3.74    &	9.51    &	21.2 \\
    19	&   3.45    &	9.44    &   3.66    &	9.47    &	21.2 \\
    21	&   3.45    &	9.36    &   3.64    &	9.38    &	21.3 \\
    23	&   3.46    &	9.45    &   3.70    &	9.60    &	21.4 \\
\hline
\end{tabular}
}
\end{table}

\subsection{Wide vs Deep} 
\label{wide_vs_deep}

To measure the effect of network widening we compare Carnelinet against its deep counterpart Citrinet. We also train a deep baseline with unified kernel size of 11. This deep model is essentially a Citrinet with all kernel sizes switched to $11$, to which we refer as \textit{Citrinet, ks=$11$}. For all models we fix the number of channels $C$ to $384$. From Table \ref{tab:ablation_wide_vs_deep} it is apparent that the deeper model with fine-tuned kernels shows a better performance, as supported by section \ref{sec:librispeech}. However, the convergence and accuracy of the deep version is quite sensitive to exact kernel size parameters and parameter unification results in poorer performance. In addition to architecture parameter stability, we also observed much better stability to change of training hyper-parameters for Carnelinet.    

\begin{table}[thb]
\centering
\caption{Wide vs Deep architectures (CarneliNet vs Citrinet). LibriSpeech, greedy WER($\%$).}
\vspace{4pt}
\label{tab:ablation_wide_vs_deep}
\scalebox{0.90}{
\begin{tabular}{ccccc} 
 \hline
 \multirow{2}{*}{\textbf{Configuration}} &
 \multicolumn{2}{c}{\textbf{dev}} &
 \multicolumn{2}{c}{\textbf{test}} \\
  & \textbf{clean} & \textbf{other} & 
    \textbf{clean} & \textbf{other}\\
 \hline
    CarneliNet-384 & 3.49 & 9.56 &	3.70 & 9.32 \\
    Citrinet-384 & 3.49 & 9.31  & 3.62 & 9.28\\
    Citrinet-384, ks=11  & 3.61 & 9.58 &	3.78 & 9.69 \\
\hline
\end{tabular}
}
\end{table}

\subsection{Tower dropout}
\label{sec:Tower_dropout}

In this section we investigate how tower dropout regularization is necessary for ensuring that ASR accuracy does not get significantly worse when model is dynamically reconfigured during inference.
For this, we compare the results of a model trained with no tower dropout (train TD = $0.0$) with the results of models that were trained with tower dropout values (train TD = $\{0.1, 0.2, 0.5\}$). We vary the number of towers removed during inference and whether that number of towers is removed from the first mega-block only, the last mega-block only, or all three blocks at once. Additionally, to quantify the trade-off between obtaining a reduced model from a larger one trained with tower dropout versus training a new model with a smaller number of towers, we provide the results for reduced models trained from scratch.

The results are shown in \autoref{tab:Tower_dropout}. If at least $1$ tower is removed, the regularized models perform much better than the baseline model trained with no tower dropout. If no towers are removed, this regularization introduces a small degradation to WER for smaller values of tower dropout. Larger values of tower dropout are only beneficial in cases when most towers are removed and introduce a considerable increase in WER when evaluated in other settings. For such extreme cases it is better to train a reduced model from scratch.

\begin{table}[thb]
\centering
\caption{Accuracy of models trained with different values of Tower Dropout (\textit{TD}) as function of the number of towers removed at inference time. We removed towers from different \textit{mega-blocks}. Models with TD are compared to the model with smaller number of towers trained \textit{from scratch}. We also provide the number of parameters (\textit{Params, M}) for the models with removed towers. CarneliNet-384, LibriSpeech, test-other, greedy WER($\%$). 
}
\vspace{6pt}
\label{tab:Tower_dropout}
\scalebox{0.85}{
\begin{tabular}{c|c|ccccc}
\hline
{\textbf{Mega}} & \multirow{2}{*}{} & \multicolumn{5}{c}{\textbf{Num. of  towers removed during inference}} \\
 \textbf{block}                          &                     & \textbf{0}   & \textbf{1}    & \textbf{2}    & \textbf{3}    & \textbf{4}     \\
                            \hline
      \multirow{6}{*}{first} &  Params, M   &   21.0	 &   19.7   &  18.8    &   17.9   &  17.0     \\ 
  \cline{2-7}
                & TD=0.0 &	            \textbf{9.32}    &	11.09   &	15.43   &	24.17   &	54.43 \\
                & TD=0.1 &	            9.70    &	\textbf{10.10}   &	\textbf{10.24}   &	11.61   &	14.84 \\
                & TD=0.2 &	            10.03   &	10.44   &	10.50   &	\textbf{11.11}   &	13.47 \\
                & TD=0.5 &	            12.35   &	12.37   &	12.56   &	12.62   &	\textbf{13.10} \\
                \cline{2-7}
                & from scratch    &	9.32    &	9.75    &	9.71    &	9.87    &	10.29 \\
                            \hline
    \multirow{6}{*}{last}   &  Params, M   &   21.0	 &   19.7   &  18.8    &   17.9   &  17.0     \\ 
  \cline{2-7}
                & TD=0.0 &	\textbf{9.32}    &	14.14   &	17.21   &	20.75   &	25.51\\
                & TD=0.1 &	9.70    &	\textbf{9.95}    &	\textbf{10.34}   &	11.16   &	12.83\\
                & TD=0.2 &	10.03   &	10.11   &	10.36   &	\textbf{11.12}   &	\textbf{11.73}\\
                & TD=0.5 &	12.35   &	12.43   &	12.53   &	12.81   &	13.03\\
                \cline{2-7}
                & from scratch    &	9.32    &	9.70    &	10.06   &	10.10   &	10.12\\
                            \hline
\multirow{6}{*}{all}   &  Params, M         &   21.1	 &   18.2   &  15.3    &   12.4   &  9.6     \\ 
  \cline{2-7}
                & TD=0.0 &	\textbf{9.32}    &	16.73   &	29.24   &	47.10   &	79.09 \\
                & TD=0.1 &	9.70    &	\textbf{10.70}   &	11.86   &	15.14   &	23.80 \\
                & TD=0.2 &	10.03   &	10.75   &	\textbf{11.53}   &	13.49   &	19.94 \\
                & TD=0.5 &	12.35   &	12.53   &	12.88   &	\textbf{13.34}   &	\textbf{14.25} \\
                 \cline{2-7}
                & from scratch    &	9.32    &	9.85    &	10.06   &	10.74   &	11.60 \\
\hline
\end{tabular}
}
\end{table}


\subsection{Weight scaling for model reconfiguration} 

The tower dropout during  training  allows us to obtain models that are robust to dynamic disabling of towers during inference. As described in \autoref{sec:architecture} if any of the sub-networks are turned off, then we adjust the summation weight factor to the rate of enabled ones to keep the mathematical expectation of the output unchanged, much like in traditional dropout.

In \autoref{tab:Tower_dropout_scaling}  we illustrate the importance of scaling the summation weight by comparing the results to the same model that does not re-scale weights. The quality of speech recognition is universally worse for the non-scaling cases. When changes are major, e.g. when removing towers from all layers, the quality degrades catastrophically without rescaling.

\begin{table}[thb]
\centering
\caption{Importance of summation weight scaling after model reconfiguration. CarneliNet-384, LibriSpeech, test-other set, greedy WER($\%$).}
\vspace{4pt}
\label{tab:Tower_dropout_scaling}
\scalebox{0.90}{
\begin{tabular}{cccccc}
\hline
{\textbf{Mega}} & \multirow{2}{*}{\textbf{Scaling?}} & \multicolumn{4}{c}{\textbf{Number of removed towers}} \\
 \textbf{block}                          &                     & \textbf{1}    & \textbf{2}    & \textbf{3}    & \textbf{4}     \\
                            \hline
      \multirow{2}{*}{first} 
                & \xmark &	            10.45   &	11.16   &	13.53   &	25.63  \\
                & \cmark &	            10.44   &	10.50   &	11.11   &	13.47  \\
                            \hline
    \multirow{2}{*}{last} 
                & \xmark &	            10.21   &	10.64   &	12.02   &	15.35 \\
                & \cmark &	            10.11   &	10.36   &	11.12   &	11.73 \\
                            \hline
\multirow{2}{*}{all}   
               & \xmark &	            11.09   &	15.11   &	43.57   &	99.55 \\
               & \cmark &	            10.75   &	11.53   &	13.49   &	19.94 \\
\hline
\end{tabular}
}
\end{table}

\subsection{Width scaling}
\label{Ablation_scaling}

The number of channels ($C$) has the single largest impact on the number of trainable parameters and on computational requirements. 
We also expect that larger models will perform better. To test this hypothesis we train models with different numbers of channels $C \in \{256, 384, 512, 768, 1024\}$. 
The obtained results are shown in \autoref{tab:Ablation_channels}. 
We observe a noticeable impact of model width on the quality of speech recognition. While extremely wide models with $C = 1024$ are able to add additional improvements to the resulting scores, there exists an upper bound to such scaling, beyond which further improvements plateau.

\begin{table}[thb]
\centering
\caption{Accuracy vs tower width ($C$ - number of channels in each tower). LibriSpeech, greedy WER($\%$).}
\label{tab:Ablation_channels}
\vspace{4pt}
\scalebox{0.90}{
\begin{tabular}{cccccc} 
 \hline
 \multirow{2}{*}{\textbf{ C}}    &
 \multicolumn{2}{c}{\textbf{dev}} &
  \multicolumn{2}{c}{\textbf{test}} & 
   {\textbf{Params,}} \\
  & \textbf{clean} & \textbf{other} & \textbf{clean}& \textbf{other} & \textbf{M}\\
 \hline
    256	    &   4.51    &	12.04   &	4.59    &	11.94   &	9.9 \\
    384	    &   3.49    &	9.56    &	3.70    &	9.32    &	21.0 \\
    512	    &   3.49    &	9.26    &	3.64    &	9.10    &	36.3 \\
    768	    &   2.96    &	8.00    &	3.24    &	7.79    &	80.8 \\
    1024    &   \textbf{2.70}    &	\textbf{7.42}    &	\textbf{3.11}    &	\textbf{7.16}    &	\textbf{141}\\
 \hline
\end{tabular}
}
\end{table}

\subsection{Depth scaling}

We analyze the dependence of the network on the number of repeated blocks $R$ in each tower. The number of blocks is kept the same across all towers, so we can vary only one parameter.
Results presented in \autoref{tab:Ablation_scaling_depth} confirm our expectations that deeper models perform better at the expense of more parameters and slower training and inference speed. 
We also observe an upper bound when scaling depth, beyond which improvements stagnate, which we find to be when depth $R$ reaches $7$, suggesting that additional architecture changes such as Residual connections between blocks are needed in this setting.

\begin{table}[thb!]
\centering
\caption{Accuracy vs Tower depth ($R$ - the number of blocks per tower). CarneliNet-384. LibriSpeech, greedy WER($\%$).}
\vspace{4pt}
\label{tab:Ablation_scaling_depth}
\scalebox{0.90}{
\begin{tabular}{cccccc} 
 \hline
 \multirow{2}{*}{\textbf{R}} &
 \multicolumn{2}{c}{\textbf{dev}} &
  \multicolumn{2}{c}{\textbf{test}} & 
   {\textbf{Params,}} \\
  & \textbf{clean} & \textbf{other} & \textbf{clean}& \textbf{other} & \textbf{M} \\
 \hline
    2   &	5.77    &	15.09   &	5.83    &	14.95   &	11.4 \\
    3   &	4.30    &	11.88   &	4.65    &	11.83   &	14.6 \\
    4   &	4.00    &	10.40   &	3.99    &	10.55   &	17.8 \\
    5   &   3.49    &   9.56    &   3.70    &   9.32    &   21.0 \\
    6   &	3.38    &	9.26    &	3.60    &	9.25    &	24.2 \\
    \textbf{7} & \textbf{3.28}  & \textbf{8.78} &\textbf{3.46} &	\textbf{8.83}    &	\textbf{27.4} \\
    8   &	3.32    &	9.17    &	3.57    &	9.22    &	30.6 \\
    9   &	3.46    &	9.46    &	3.68    &	9.37    &	33.8 \\
    10  &	3.46    &	9.32    &	3.72    &	9.17    &	36.9 \\
 \hline
\end{tabular}
}
\end{table}

\subsection{Parallel QuartzNet}
\label{sec:ablation_parallel_quartznet}

In order to demonstrate that a mixture model can be successfully applied not only to the Citrinet model but to various models in general, we perform a similar transformation to QuartzNet-15x5 \cite{kriman2020quartznet}. Since QuartzNet contains blocks with different kernel sizes and number of filters, the grouping of computational sub-networks can be done in multiple ways. For example, one can choose to put kernels with the same number of filters in the same mega-block.  However, we found that grouping by the same kernel size gives the best results. We trained this model (\textit{Parallel QuartzNet}) on LibriSpeech, with most hyper-parameters borrowed from the original QuartzNet in the paper.  Due to the increased width, we added  dropout of $0.1$, performed strong augmentation with SpecAugment \cite{park2019}, and raised initial LR to $0.1$. 

Table \ref{tab:ablation_parallel_quartznet} shows the results of the trained models evaluated on LibriSpeech. While there is a gap of $1.6\%$ in WER score, we note that the parallel version still successfully trains and comes with benefits of dynamic reconfiguration during inference without retraining.
This demonstrates that the process of constructing a mixture model can be performed for different sub-networks and indicates an opportunity for experiments on different architectures.

\begin{table}[thb]
\centering
\caption{Deep \textit{QuartzNet-15x5} vs \textit{Parallel QuartzNet} with blocks grouped by kernel size. LibriSpeech, 400 epochs, greedy WER($\%$).}
\vspace{4pt}
\label{tab:ablation_parallel_quartznet}
\scalebox{0.90}{
\begin{tabular}{ccccc} 
 \hline
 \multirow{2}{*}{\textbf{Configuration}} &
 \multicolumn{2}{c}{\textbf{dev}} &
 \multicolumn{2}{c}{\textbf{test}} \\
  & \textbf{clean} & \textbf{other} & 
    \textbf{clean} & \textbf{other}\\
 \hline
    QuartzNet & 3.83 & 11.08 &	3.90 & 11.28 \\
    \hline
    Parallel QuartzNet &  5.17 &	12.61 & 	5.13 &	12.84  \\
\hline
\end{tabular}
}
\end{table}

\section{Conclusions}
\label{sec:conclusions}

In this work we discussed the implementation and advantages of a parallel mixture of shallow networks for ASR. We illustrated the effectiveness of such composition with a novel CTC-based architecture CarneliNet composed of three mega-blocks with parallel sub-networks based on 1D depthwise-separable convolutions. We evaluated the model on LibriSpeech, MLS and AISHELL-2 datasets and achieved close to state-of-the-art results for non-autoregressive CTC-based models on LibriSpeech and MLS and state-of-the-art results for non-autoregressive models on AISHELL-2. We demonstrated how such a model with parallel sub-networks can be dynamically reconfigured without retraining to adjust to computation requirements. We believe that this approach can be transferred to additional architectures, as supported by the Parallel QuartzNet experiments.

The models and training recipes have been released in the NeMo toolkit~\cite{kuchaiev2019nemo}.\footnote{\url{https://github.com/NVIDIA/NeMo}} 

\section{Acknowledgments}
The authors thank Elena Rastorgueva, Jocelyn Huang for extensive review of the manuscript and the whole NVIDIA AI Applications team for their helpful feedback and review.

\bibliography{refs}

\clearpage 



\end{document}